\documentclass[twocolumn]{aastex63}
\usepackage{amsmath}
\usepackage{color}
\usepackage{graphicx}
\usepackage{subfigure}
\usepackage{float}
\usepackage{textcomp}
\usepackage[T1]{fontenc}
\usepackage{ae,aecompl}
\usepackage{amssymb}
\usepackage{booktabs}

\begin{document}

\title{Effects of vertical advection on multimessenger signatures of black hole neutrino-dominated accretion flows in compact binary coalescences}

\correspondingauthor{Tong Liu}
\email{tongliu@xmu.edu.cn}

\author[0000-0003-4041-7700]{Bing-Guang Chen}
\affiliation{Department of Astronomy, Xiamen University, Xiamen, Fujian 361005, China}

\author[0000-0001-8678-6291]{Tong Liu}
\affiliation{Department of Astronomy, Xiamen University, Xiamen, Fujian 361005, China}

\author[0000-0002-1768-0773]{Yan-Qing Qi}
\affiliation{Department of Astronomy, Xiamen University, Xiamen, Fujian 361005, China}

\author[0000-0002-4448-0849]{Bao-Quan Huang}
\affiliation{Department of Astronomy, Xiamen University, Xiamen, Fujian 361005, China}

\author[0000-0002-9130-2586]{Yun-Feng Wei}
\affiliation{Department of Astronomy, Xiamen University, Xiamen, Fujian 361005, China}

\author[0000-0002-5839-6744]{Tuan Yi}
\affiliation{Department of Astronomy, Xiamen University, Xiamen, Fujian 361005, China}

\author[0000-0003-3137-1851]{Wei-Min Gu}
\affiliation{Department of Astronomy, Xiamen University, Xiamen, Fujian 361005, China}

\author[0000-0002-0279-417X]{Li Xue}
\affiliation{Department of Astronomy, Xiamen University, Xiamen, Fujian 361005, China}

\begin{abstract}
In the coalescence events of binary neutron star (NS) or a black hole (BH) and an NS, a BH hyperaccretion disk might be eventually formed. At very high mass accretion rates, MeV neutrinos will be emitted from this disk, which is called a neutrino-dominated accretion flow (NDAF). Neutrino annihilation in the space out of the disk is energetic enough to launch ultrarelativistic jets to power gamma-ray bursts. Moreover, vertical advection might exist in NDAFs, which can generate the magnetic buoyancy bubbles to release gamma-ray photons. In this paper, we visit the effects of the vertical advection in NDAFs on the disk structure and gamma-ray and neutrino luminosities for different accretion rates. Then we study the anisotropic emission of kilonovae and the following gravitational waves (GWs) driven by the gamma-ray photons and neutrinos from NDAFs. Comparing NDAFs without vertical advection, the neutrino luminosity and GW strains slightly decrease for the case with vertical advection, and the kilonovae will be brightened by the injected gamma-ray photons. The future joint multimessenger observations might distinguish whether the vertical advection exists in NDAFs or not after compact binary coalescences.
\end{abstract}


\section{Introduction}
	
Short-duration gamma-ray bursts (SGRBs) commonly occur in the scenario of compact binary coalescences, i.e., binary neutron star (NS) or a black hole (BH) and an NS. When the binary gets into gravitational fields of each other and spiral towards one another, it will radiate gravitational waves \citep[GWs, e.g.,][]{Cutler1994,Sathyaprakash2009,Baiotti2017}. After coalescence events, optical/near-infrared emission from the radioactive decay of rapid neutron captured ($r$-process) elements are produced by the neutron-rich ejecta during the merger and postmerger phases \citep[e.g.,][]{Lattimer1974,Lattimer1976,Symbalisty1982,Li1998,Kasen2013,Just2015,Metzger2019,Nakar2020,Cowan2021}. These transient events were named as ``kilonovae'' because their luminosity was approximately 1,000 times brighter than typical novae and its emission timescale would last for several days or even longer \citep[e.g.,][]{Metzger2010,Metzger2017,Metzger2019}. Besides, the remnants are expected to emit large amounts of MeV neutrinos that is because the violent collision heats up the material which goes through rapid decompression from the debris of compact object \citep[e.g.,][]{Sekiguchi2011,Kyutoku2018}. Meanwhile, the newborn BH constantly devours matters and thus produces ultrarelativistic jets from bipolar, which would be detected as SGRBs if they point toward the Earth \citep[e.g.,][]{Paczynski1986,Narayan1992,Popham1999,Liu2017a}.

The successful observations to GW 170817 by the advanced Laser Interferometer Gravitational-Wave Observatory (aLIGO)/Virgo \citep{Abbott2017a} and its accompanied GRB 170817A \citep[e.g.,][]{Abbott2017b} has indicated the beginning of a new era of multimessenger astronomy; and about 10 h after merger event, the electromagnetic counterpart, kilonova AT 2017gfo, was detected \citep[e.g.,][]{Abbott2017b,Andreoni2017,Arcavi2017,Coulter2017}, which was indubitably another landmark of multimessenger signals. Its light curve cannot be explained in the $r$-process depended kilonova model by using only one single set of parameters. Instead, the different mass, velocity, morphology, and opacity of the ejecta all need to be taken into consideration \citep[e.g.,][]{Cowperthwaite2017,Perego2017,Tanaka2017,Villar2017,Kawaguchi2018,Wu2019}. Following the observation of GW 170817, IceCube, ANTARES, Super-Kamiokande \citep{Abe2018}, and the Pierre Auger Observatory \citep{Albert2017} all attempted to detect the accompanying high-energy neutrinos, but it is rather unfortunate that none of them received significant neutrinos from GW 170817. After that, on January 5, 2020, the LIGO/Virgo detected the first BH-NS merger event, GW 200105, and 10 days later, the second BH-NS merger event, GW 200115, was also discovered \citep{Abbott2021}. Both accretion and ejection of material from a BH produce electromagnetic radiation, which was not observed in either of these events \citep[e.g.,][]{Zhu2021,Qi2022}. It is possible that the NS is not disintegrated but is swallowed intactly by the BH and cannot produce noticeable electromagnetic counterpart.

A stellar-mass BH surrounded by a hyperaccretion disk will be formed in the center of compact binary coalescences or massive collapsars. For the very high accretion rates ($10^{-3}\ M_{\odot}\ \mathrm{s^{-1}} \lesssim \dot{M} \lesssim 10\ M_{\odot}\ \mathrm{s^{-1}}$), such disk has extremely density ($\rho \sim 10^{10}-10^{13} \mathrm{~g} \mathrm{~cm}^{-3}$) and temperature ($T \sim 10^{10}-10^{11}$ K). In the inner region of the disk, neutrino cooling will be dominated to balance the viscous heating. This accretion disks is named as neutrino-dominated accretion flows \citep[NDAFs; for reviews, see][]{Liu2017a,Zhang2018}. It is first proposed by \cite{Popham1999} and explains the energy supply of gamma-ray bursts (GRBs) in terms of neutrino annihilation in the space out of the disk. Subsequently, the NDAF model was widely studied \citep[e.g.,][]{Narayan2001,DiMatteo2002,Kohri&Mineshige2002,Kohri2005,Lee2005,Gu2006,Chen&Beloborodov2007,Kawanaka&Mineshige2007,Liu2007,Lei2009,Zalamea2011,Kawanaka2013,Xue2013}. Moreover, three-dimensional radiation magneto-hydrodynamical simulation by \citet{Jiang2014} and \citet{Jiang2019} revealed the vertical advection process. They found that it caused by magnetic buoyancy transports allows a significant fraction of gamma-ray photons to escape from the surface of the disk before being advected into the BH. The vertical advection in NDAFs was investigated by \cite{Yi2017}, and their results shows that the effect can change the structure and increase the luminosity of NDAFs. We consider that the abundant gamma-ray photons induced by the vertical advection should be injected into the ejecta to brighten the kilonovae and also affect the emission of neutrinos and therefore GWs triggered by NDAFs.

In this paper, we investigate the NDAF with vertical advection around a fast-rotating BH (as shown in Figure \ref{fig:disk}) and the effects of the vertical advection on the disk structure, kilonovae, and neutrino and GW emissions. The paper is structured as follows. In Section 2, we model the NDAFs with vertical advection, the anisotropic kilonovae induced by the ejecta and gamma-ray photons together, the spectra of electron antineutrino from NDAFs by considering the general relativistic effects, and GW emission leading to the anisotropic neutrinos from NDAFs. The main results are shown in Section 3. A brief summary is made in Section 4.

\section{Model}
\subsection{NDAFs with vertical advection}

\begin{figure}
\centering
\includegraphics[width=0.4\textwidth]{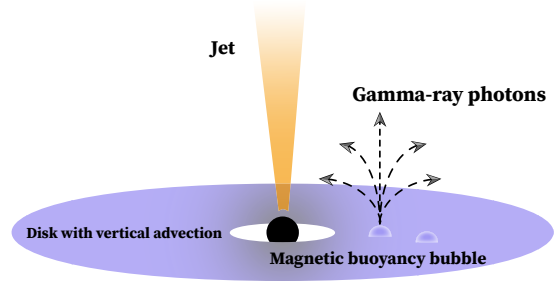}
\caption{Schematic diagram of the NDAF with vertical advection.}
\label{fig:disk}
\end{figure}

Considering that BHs accrete gas constantly, this causes it to accumulate angular momentum, and in most GRBs engine candidates, central BHs are rotating rapidly, we adopt the relativistic correction factors \citep{Riffert1995} to describe the Kerr BHs, i.e.,
\begin{align}
\label{NDAFfirst} A&=1-\frac{2 G M_{\mathrm{BH}}}{c^{2} r}+\left(\frac{G M_{\mathrm{BH}} a_{*}}{c^{2} r}\right)^{2},\\
B&=1-\frac{3 G M_{\mathrm{BH}}}{c^{2} r}+2 a_{*}\left(\frac{G M_{\mathrm{BH}}}{c^{2} r}\right)^{3 / 2},\\	
C&=1-4 a_{*}\left(\frac{G M_{\mathrm{BH}}}{c^{2} r}\right)^{3 / 2}+3\left(\frac{G M_{\mathrm{BH}} a_{*}}{c^{2} r}\right)^{2},\\	
D&=\int_{r_{\rm m s}}^{r} \frac{\frac{x^{2} c^{4}}{8 G^{2}}-\frac{3 x M_{\mathrm{BH}} c^{2}}{4 G}+\sqrt{\frac{a_{*}^{2} M_{\mathrm{BH}}^{3} c^{2} x}{G}}-\frac{3 a_{*}^{2} M_{\mathrm{BH}}^{2}}{8}}{\frac{\sqrt{r x}}{4}\left(\frac{x^{2} c^{4}}{G^{2}}-\frac{3 x M_{\mathrm{BH}} c^{2}}{G}+2 \sqrt{\frac{a_{*}^{2} M_{\mathrm{BH}}^{3} c^{2} x}{G}}\right)}dx,
\end{align}
where $r$ and $r_{\rm ms}$ are the disk radius and the inner boundary of the disk, respectively. In our model, we consider a fast-rotating stellar-mass BH with the mass $M_{\mathrm{BH}} =3~ M_{\odot}$ and the dimensionless spin parameter $a_* = 0.9$.

The kinematic viscosity is written as
\begin{equation}
\nu=\alpha \frac{c_{s}^{2}}{\Omega_{\rm K}},
\end{equation}
where $\alpha = 0.1$ is the viscous parameter of the disk, $c_{s}=\sqrt{P / \rho}$ is the isothermal sound speed with $P$ being the pressure and $\rho$ the density, and $\Omega_{\rm K}=\left(G M_{\rm BH} / r^{3}\right)^{1 / 2}$ is the Keplerian angular velocity. In the Kerr metric, the basically dynamic equations of NDAFs are given as follows \citep[e.g.,][]{Liu2010,Liu2017a}
\begin{equation}
\dot{M}=-4 \pi r v_{r} \rho H ,
\end{equation}
and
\begin{equation}
\dot{M} \frac{D}{A}=4 \pi \nu \rho H \sqrt{\frac{A}{B C}}\ ,
\end{equation}
where $H = \sqrt{{P r^{3}}/{\rho G M_{\rm BH}}} \sqrt{{B}/{C}}$ is the half-thickness of the disk and $v_r$ is the radial velocity of the accreted gas. Here two typical accretion rates $\dot{M} = 0.1$ and $1\ \ M_{\odot}\ \mathrm{s^{-1}}$ are considered.

The total pressure consists of four items, i.e., radiation pressure, gas pressure, electron degeneracy pressure, and neutrino pressure. Thus the equation of state is expressed as \citep[e.g.,][]{Liu2010}:
\begin{equation}
\begin{aligned}
P=& \frac{11}{12} a T^{4}+\frac{\rho k T}{m_\mathrm{p}}\left(\frac{1+3 X_{\mathrm{nuc}}}{4}\right) \\ &+\frac{2 \pi h c}{3}\left(\frac{3}{8 \pi m_\mathrm{p}}\right)^{4 / 3}\left(\frac{\rho}{\mu_\mathrm{e}}\right)^{4 / 3}+\frac{u_{\nu}}{3},
\end{aligned}
\end{equation}
where $a$ is the radiation constant, $m_\mathrm{p}$ is the proton rest mass, $k$ is the Boltzmann constant, $h$ is the Planck constant, $\mu_\mathrm{e}$ is the electron chemical potential, and $X_{\mathrm{nuc}} \simeq$ $34.8 \rho_{10}^{-3 / 4} T_{11}^{9 / 8} \exp \left(-0.61 / T_{11}\right)$ is the mass fraction of free nucleons with $T_{11}=T / (10^{11} \mathrm{~K})$ and $\rho_{10}=\rho / (10^{10} \mathrm{~g} \mathrm{~cm}^{-3})$ \citep[e.g.,][]{Gu2006,Liu2007}. In addition, the neutrino energy density $u_\mathrm{\nu}$ is given by \citep[e.g.,][]{Popham1995,Liu2007}
\begin{equation}
u_{\nu}=(7 / 8) a T^{4} \sum \frac{\tau_{\nu_{j}} / 2+1 / \sqrt{3}}{\tau_{\nu_{j}} / 2+1 / \sqrt{3}+1 /\left(3 \tau_{a, \nu_{j}}\right)},
\end{equation}
where $\tau_{\nu_j}=\tau_{\mathrm{a}, j}+\tau_{\mathrm{s}, j}$ is the sum of the total absorptive and scattering optical depths for each neutrino flavor $\left(\nu_{e}, \nu_{\mu}, \nu_{\tau}\right)$ \cite[e.g.,][]{DiMatteo2002,Gu2006,Liu2010}.

The energy equation is written as
\begin{equation}
Q_\mathrm{vis}^{+} = Q_\mathrm{adv}^{-} + Q_\mathrm{\nu}^{-} + Q_\mathrm{photo}^{-} + Q_\mathrm{z}^{-}.
\end{equation}
The above equation illustrates the balance between heating because of viscous dissipation $Q_\mathrm{vis}^{+}$ and cooling due to advection $Q_\mathrm{adv}^{-}$, neutrino losses $Q_\mathrm{\nu}^{-}$, photodisintegration $Q_\mathrm{photo}^{-}$, and vertical advection $Q_\mathrm{z}^{-}$ in turn.

The heating rate is expressed as \citep[e.g.,][]{Liu2010}
\begin{equation}
Q_{\mathrm{vis}}^{+}=\frac{3 G M \dot{M}}{8 \pi r^{3}} \frac{D}{B},
\end{equation}
and we take the advective cooling term in \cite{Liu2007}, which is written as
\begin{equation}
Q_{\mathrm{adv}}^{-} \simeq v_{r} \frac{H}{r}\left(\frac{11}{3} a T^{4}+\frac{3}{2} \frac{\rho k T}{m_{\mathrm{p}}} \frac{1+X_{\mathrm{nuc}}}{4}+\frac{4 u_{\nu}}{3}\right).
\end{equation}
We adopt a bridging formula for calculating the neutrino transport, i.e.,
\begin{equation}
Q_{\nu}^{-}=\sum \frac{\left(7 / 8 \sigma T^{4}\right)}{(3 / 4)\left(\tau_{\nu_{j}} / 2+1 / \sqrt{3}+1 /\left(3 \tau_{a, \nu_{j}}\right)\right)}.
\end{equation}
Moreover, the rate of the cooling due to photodissociation is always ignored because it is much less than the neutrino cooling rate in the inner disk \citep[e.g.,][]{Liu2007}.

The vertical advection effect leads to the magnetic buoyancy bubbles, carrying the gamma-ray photons and floating up from the equatorial plane of the disk. We introduce a new cooling term $Q_\mathrm{z}^{-}$ to describe this process. Figure \ref{fig:disk} provides a schematic description of an NDAF with vertical advection effect.

First, we consider that the averaged velocity of the vertical advection process, $\overline{V_{z}}$, can be estimated by
\begin{equation}\label{NDAFlast}
\overline{V_z} = \lambda c_s,
\end{equation}
where $\lambda$ is a dimensionless parameter, and we take $\lambda=0.1$ by referring to the typical vertical velocity given by the numerical simulations in \cite{Jiang2014}. Second, based on the mixing-length theory \citep{Prandtl1925}, a turbulent eddy viscosity is related to the mixing length $\ell$ and the gradient of mean velocity, i.e., $\nu = \ell^2 \mid\partial \bar{u} / \partial z\mid$. Thus, by considering Equations (5) and (14), we can roughly estimate the mixing length of the bubbles $\ell \sim H$, suggesting that these bubbles can reach the disk surface. Third, the optical depth of the bubbles is mainly determined by the vertical distribution of the disk density. In order to hold the dynamical equilibrium at the surface of bubbles, the bubbles should expand and keep the lower density therein when they float up. Finally, they will mix with the circumstance at the disk surface. The optical depth over there is thin enough to release photons. Moreover, the process of the bubble expansion is nonadiabatic, because photons could be produced at any height, and the matter is extremely dense and optically thick for them, so these photons tend to inject into the existing bubbles or generate new bubbles, where the density is lower than that of NDAFs. In absence of detailed modeling of the disk vertical structure, we utilize physical quantities at the disk equatorial plane to characterize the cooling process of vertical advection in NDAFs for simplicity.

Based on the above analyses, we adopt the formula,
\begin{equation}
Q_\mathrm{z}^{-}=\overline{V_{z}}\left(u_{\mathrm{ph}}+u_{\nu}+u_{\mathrm{gas}}\right),
\end{equation}
where $u_{\mathrm{ph}}$, $u_{\nu}$, and $u_{\mathrm{gas}}$ are the energy density of photons, neutrinos, and gases, respectively \citep{Yi2017}. In our calculation, the third term $\overline{V_z} u_{gas}$ is ignored because the amount of gas escaped through the magnetic buoyancy is tiny compared to the other terms. The vertical advection term describes the released photons and neutrinos due to the magnetic buoyancy, which can dominate over the normal diffusion process.

Although the roughly estimated mixing length of the bubbles approaches the half disk thickness, one can notice that not all of the bubbles from any radius of the disk could reach the disk surface in simulations (see Figures 9 and 14 of \cite{Jiang2014}). Moreover, the yields of the photons will decrease once the place of production is deviation from the equatorial plane of the disk. Thus we consider that this description denotes the upper limit of the cooling rate due to the vertical advection, and then the gamma-ray photons released from bubbles as the new sources of the energy injection, their contribution to brighten up the kilonovae might be exaggerated to a certain extent.

\subsection{Kilonovae}

As first suggested by \cite{Lattimer1974}, the material that ejected from compact binary mergers is a favorable place for the generation of heavy elements through $r$-process nucleosynthesis. These extremely erratic heavy nuclei promptly decay to power the so-called ``kilonovae'' \citep[e.g.,][and references therein]{Li1998, Metzger2010,Kasen2013,Tanaka2013,Liu2017a,Metzger2017,Metzger2019}. The interpretation works on AT 2017gfo indicate that a single component model of kilonova is not sufficient to characterize the observations in multi-bands, so two or more components are required \citep[e.g.,][]{Cowperthwaite2017, Tanaka2017, Tanvir2017}. It is widely accepted that kilonova emission covers multiple phases, from the early optical and ultraviolet band which is known as the ``blue'' component (BC) to the later infrared band which called ``red'' component (RC). The occurrence of this phenomenon can be attributed to the fact that the opacity $\kappa$ of the merger ejecta may not a single value, but depends on the composition of the ejecta material, which is caused by the different proportions of lanthanides and actinides in it. When the ejected material undergoes more thoroughly $r$-process nucleosynthesis, it will generate abundant lanthanides and actinides, thus it can explain the RC because lanthanides can reach to a high opacity $\kappa \gtrsim \mathrm{10\ cm^2\ g^{-1}}$, much higher than substances composed of iron group elements \citep[e.g.,][]{Tanaka2013,Barnes2013}; but if the ejecta experiences a partially $r$-process nucleosynthesis, it will prevent it from the formation of lanthanide elements, so that the BC appears, which has a lower typical opacity $\mathrm{\kappa \sim 0.1\ cm^2\ g^{-1}}$ \citep[e.g.,][]{Cowperthwaite2017}.

The materials from compact binary merger could be ejected from the whole solid angle, like a thick crust of spheres with a central engine in the middle, but this shell geometry is not necessarily uniform distribution, that is, the ejecta no need to be isotropic. In fact, the dynamic ejecta produced by the BH-NS merger is primarily along the equatorial plane, and looks like a crescent-shaped \citep[e.g.,][]{Kyutoku2013,Kyutoku2015,Kawaguchi2016,Zhu2020,Qi2022}. The ejected mass varies from ${10^{-3}}$ to ${10^{-2}~\ M_{\odot}}$ and a relativistic velocity that accelerates with ejecta diffusion from 0.1 to 0.4 $c$. Particularly, in the equatorial direction, where the most ejection happens, neutron-rich ejecta can produce large lanthanide contents. This part of the ejecta is primarily produced via the tidal tail which has a high opacity thus reddens the kilonovae, i.e., the RC. On the other hand, in the polar direction, where the ejecta might be injected by the neutrino-driven disk winds or outflows, the electron abundance is high therefore the $r$-process is restricted and elements with atomic mass number $\mathrm{A \ge 130}$ cannot be readily synthesized. The kilonovae in this direction will look more ``blue'', i.e., the BC. Based on the above considerations, the differences in the geometry and opacity of the ejecta in each angle make it necessary to be explained by using an anisotropic kilonova emission model.

\begin{figure}
\centering
\includegraphics[scale=0.18]{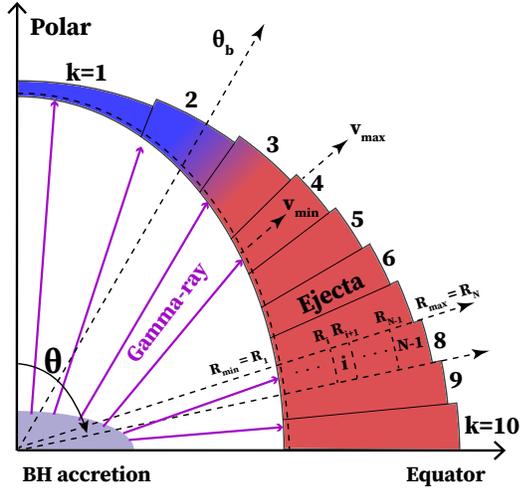}
\caption{Schematic diagram of coalescence ejecta and injected gamma-ray photons.}
\label{fig:kilonova}
\end{figure}

In addition to the radioactive decay of the ejecta, there are a few others potentially important energy outputs such as the neutrino radiation from the accretion disk is expected to heat the surroundings, creating the disk winds \citep[e.g.,][]{McLaughlin2005,Metzger2008,Dessart2009}, and winds from magnetized NS \citep[e.g.,][]{Yu2013,Yu2018,Metzger2014,Ren2019}. The interaction between relativistic jets and ejecta can also cause shock heating \citep[e.g.,][]{Bucciantini2012,Gottlieb2018,Piro2018}. Besides, a large amount of works had taken the accretion disk outflows into account \citep[e.g.,][]{Just2015,Wu2016,Perego2017,Siegel2017,Song2018,Qi2022}, which is another important source of mass and energy.

In our model, with the presence of vertical advection mechanism, the gamma-ray photons that have escaped from the magnetic buoyancy bubbles on the disk carry enormous energy that will trigger brighter kilonovae, which could have a further impact on the observations. As we mentioned before, after the coalescence event, the formation of a fast-rotating central BH largely affects the gravitational field in its vicinity including the direction of photon emission from the disk. These photons will move along the geodesic of the Kerr metric, that is, this will intensify their anisotropic emission. In order to calculate these trajectories of photons, we use ray-tracing method \citep{Fanton1997} and apply it also to the calculation of neutrinos (see Section 2.3); and this radiant energy is supposedly injected into the nearest material (normal incidence is assumed), which is the first shell layer of the ejecta. As shown in Figure \ref{fig:kilonova}, we divide it into $N (\gg 1)$ layers, and use the subscript to denote the mass layers $i=1,2, \cdots, N$, where $i = 1$ and $N$ represent the innermost and outermost layers, respectively. Each one has a different expansion speed $v_i$, the first layer has the lowest speed ${v_{1}=v_{\mathrm{min} }}$ and the outermost layer has the maximum ${v_N = v_{\mathrm{max}}}$,  based on a simplified radiation transfer model. We set the values of $v$ from 0.1 to 0.3 $c$. \citep[e.g.,][]{Kasen2010,Metzger2019,Qi2022}.

\begin{figure*}
\centering
\includegraphics[width=0.49\textwidth]{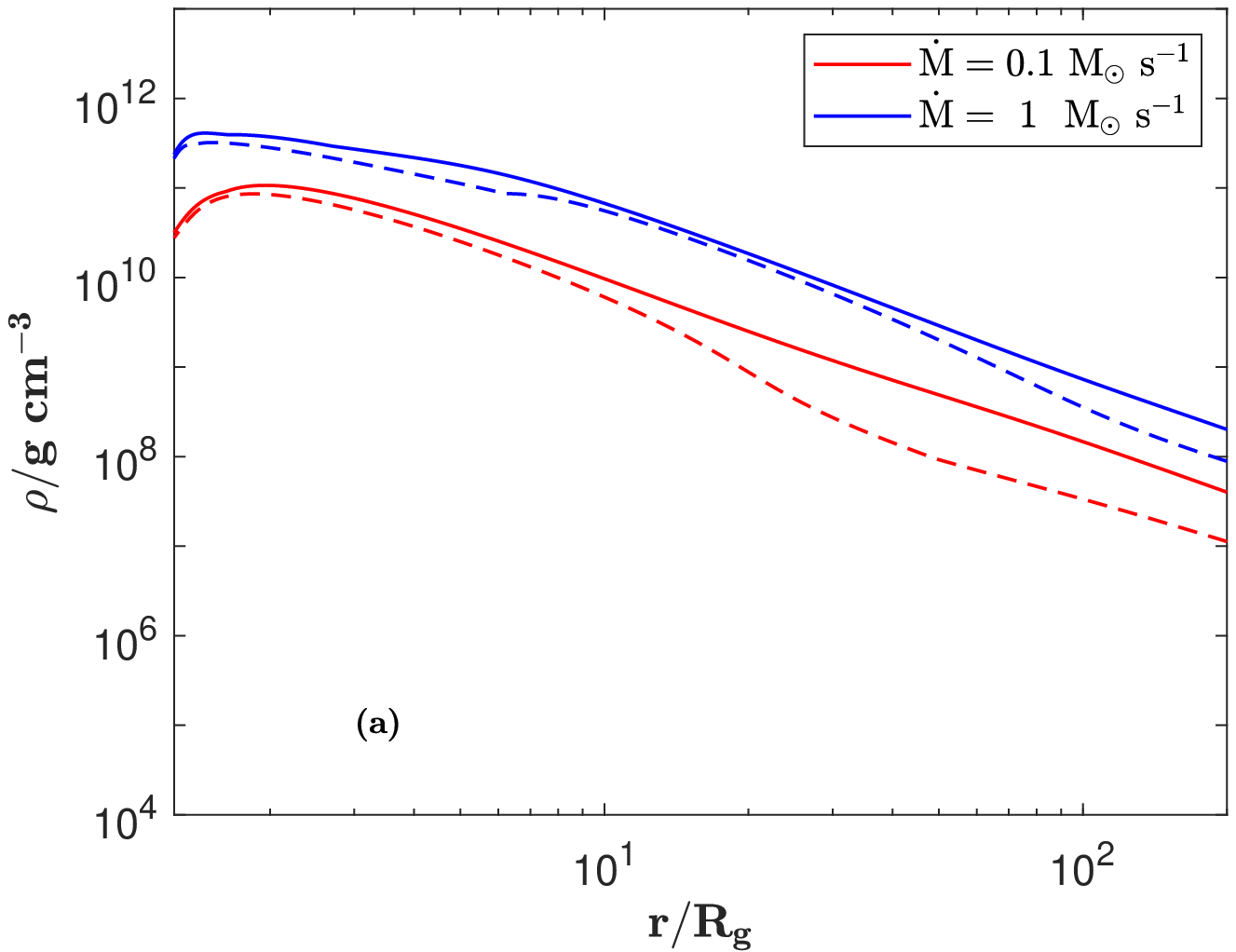}
\includegraphics[width=0.49\textwidth]{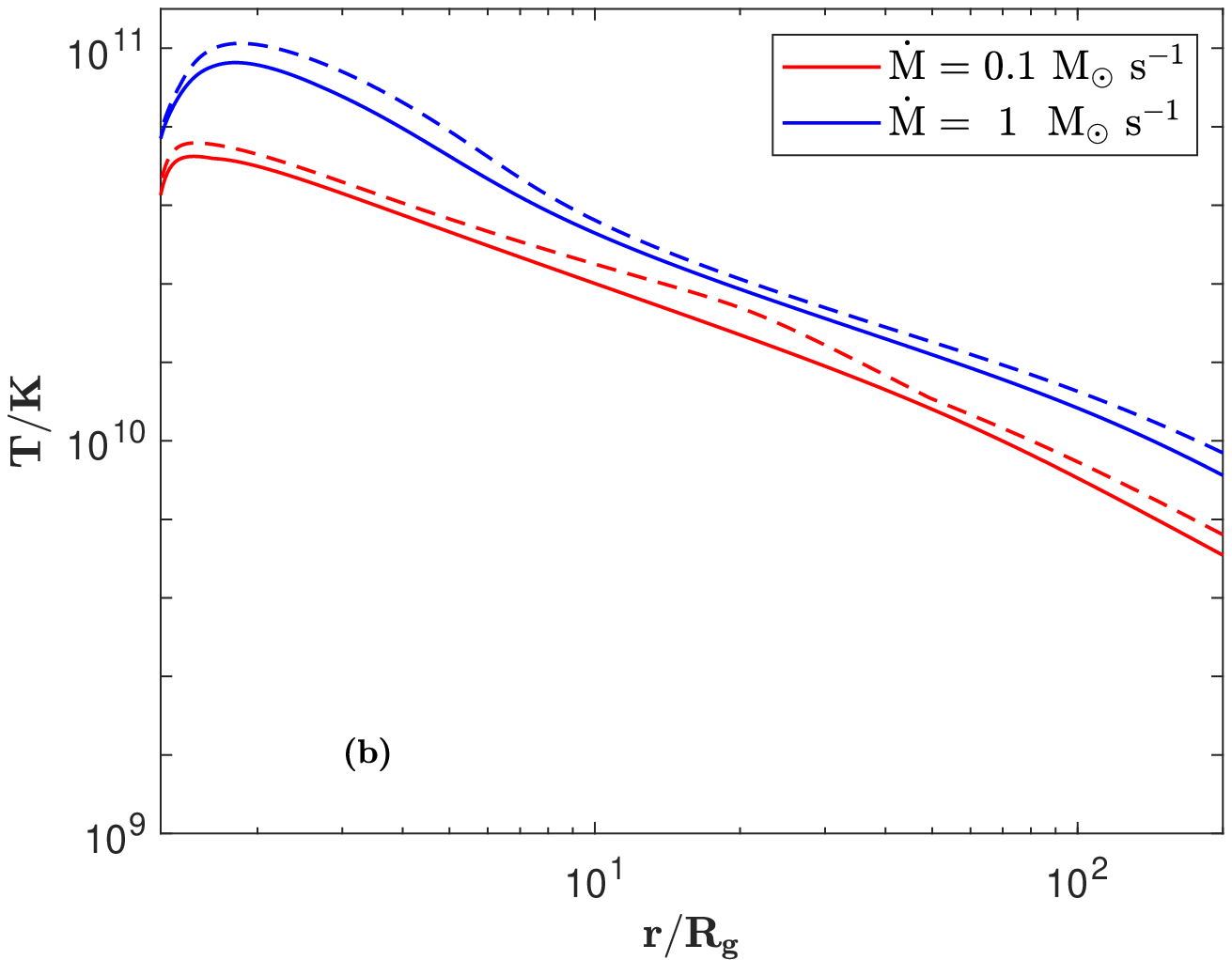}
\caption{Radial profiles of density and temperature of NDAFs with vertical advection for mass accretion rates $\dot{M} = 0.1$ and $1 \ M_{\odot}\ \mathrm{s^{-1}}$. The solid and dashed lines represent the results with and without vertical advection, respectively. Note that $R_g = 2GM_{\rm BH}/c^2$ is the Schwarzschild radius.}
\label{fig:NDAFsolution}
\vspace{0.2in}
\end{figure*}

At a time $t$, $R_i(t) = v_i t$ represents the location of the $i$th layer, so the maximum and minimum position of the ejecta can be easily calculated by $R_{\mathrm{max} }(t)=v_{\mathrm{max} } t$ and $R_{\mathrm{min} }(t)=v_{\mathrm{min}} t$. The density distribution is taken as \citep[e.g.,][]{Nagakura2014}
\begin{equation}
	\rho_{\mathrm{ej},k}=\frac{(3-\delta) m_{\mathrm{ej}, k}}{\Delta\Omega_k R_{\max }^{3}}\left[1-\left(\frac{R_{\min }}{R_{\max }}\right)^{3-\delta}\right]^{-1}\left(\frac{R}{R_{\max }}\right)^{-\delta},
\end{equation}
where $\delta$ is the power-law index of the density distribution between 1 and 3. The ejecta is supposed to be symmetric along the equatorial axis; and because of its crescent-shaped distribution, we manually partition $k$ ($k=1,2,...,10$) discrete clumps in equivalent solid angles $\Delta\Omega_k$ from the polar axis to the equatorial plane, with an increasing mass for each block as shown in Figure 2. The overall mass of the ejecta should be around several times of $0.01\ M_{\odot}$ \citep{Arnett1980}. Here we set the total mass of the ejecta, $M_{\mathrm{ej}}= 0.04\ M_{\odot}$, and the mass distribution of the ejecta meets $\sim {\theta}^2$.

At a solid angle $\Delta \Omega_k$, the emission of the ejecta is correlated to the thermal energy $E_i$ for which the evolution can be described as \citep[e.g.,][]{Ren2019}
\begin{equation}\label{N2014}
\begin{split}	
\frac{d E_{i}}{d t}=\left(1-e^{-\Delta \tau_i}\right) e^{\tau_i-\tau_{\mathrm{tot}}} L_{\mathrm{ph}}
+m_{i} \dot{q}_{\mathrm{r}} \eta_{\mathrm{th}}-\frac{E_{i}}{R_{i}} \frac{d R_{i}}{d t}-L_{i},\  \\ \mathrm { for }\ i=1,2,\ldots, N,
\end{split}
\end{equation}
where $L_\mathrm{ph}$ is the injected power carried by a gamma-ray photons, $m_i$  and $L_i$ are the mass and radiation luminosity of the $i$th layer, $\dot{q}_{\mathrm{r}}$ is the radioactive decay power per unit mass, and $\eta_{\mathrm{th}}$ is the thermalization efficiency. By considering the optical depth, on the right-hand side of Equation (\ref{N2014}), these four terms are the injected energy absorbed by the $i$th layer, radioactive decay in each layer, the cooling attributed to adiabatic expansion, and radiation, respectively. Here $\tau_i=\sum_i^{N-1} \Delta \tau_i $ is the optical depth of the $i$th layer to $N-1$ and the total optical depth is the aggregation of all layers $\tau_{\rm tot}= {\sum\limits_{i=1}^{N-1}} \Delta\tau _{i}$, where the optical depth of the $i$th layer is described as $\Delta \tau _{i} = \int_{R_{i}}^{R_{i+1}} \kappa \rho_{\rm ej} d R$. It should be mentioned that when the gamma-ray photons reach the innermost layer, they can be easily absorbed, and afterwards, as the optical depth increases dramatically, they will have a negligible effect on the layers that follow.

The radioactive power per unit masses derived from elaborate nucleosynthesis calculations \citep[e.g.,][]{Korobkin2012}
\begin{equation}
\dot{q}_{\mathrm{r}}=4 \times 10^{18}\left[\frac{1}{2}-\frac{1}{\pi} \arctan \left(\frac{t-t_{0}}{\sigma}\right)\right]^{1.3} \mathrm{erg}\  \mathrm{s}^{-1} \mathrm{~g}^{-1},
\end{equation}
where $\sigma = 0.11$ s, $t_0 = 1.3$ s, and thermalization efficiency can be written as \citep[e.g.,][]{Barnes2016}
\begin{equation}
\eta_{\text {th }}=0.36\left[\exp \left(-0.56 t_{\text {day }}\right)+\frac{\ln \left(1+0.34 t_{\text {day }}^{0.74}\right)}{0.34 t_{\text {day }}^{0.74}}\right]
\end{equation}
with $t_{\mathrm{day}}=t / 1 \mathrm{day}$. This equation is satisfied in the case of $m_{\rm ej} = 0.01~M_\odot$, $v = 0.1 c$, with random magnetic fields. In order to ensure that our calculations are compatible with this equation, as mentioned above, we divided the mass into 10 clumps, where the mass of four blocks near the equator is in the range of $5 \times 10^{-3}$ to 0.01 $M_{\odot}$.

$L_i$ is the observed radiation luminosity of the $i$th layer, which can be calculated by \citep[e.g.,][]{Yu2018,Ren2019}
\begin{equation}
L_{i}=\frac{E_{i}}{\max (t_{\mathrm{d}, i}, t_{\mathrm{lc}, i})},
\end{equation}
where the light crossing time $t_{\mathrm{lc}, i}=R_{i} / c$ gives the time limit, and $t_{\mathrm{d}, i}$ represents the radiation diffusion timescale during which the thermal heat can escape from the entire ejecta
\begin{equation}
t_{\mathrm{d}, i}=\frac{3 \kappa}{\Delta \Omega_k R_{i} c} \sum_{n=i}^{N-1} m_{n}.
\end{equation}
The form of $\max (t_{\mathrm{d}, i}, t_{\mathrm{lc}, i})$ guarantees the causality, especially in the optically thin layer which is close to the outermost shell.

By summarizing the contributions of each layer and all $\Delta \Omega_k$, the total bolometric luminosity of the merger ejecta can be obtained by
\begin{equation}
L_{\mathrm{bol}}=\sum_{i} L_{i}.
\end{equation}

Assuming that the spectrum is always blackbody radiation and is beaming from the photosphere $R_{\mathrm{ph}}$. Then, the effective temperature of the kilonova emission can be calculated as follows
\begin{equation}
T_{\mathrm{eff}}=\left(\frac{L_{\mathrm{bol}}}{\sigma_{\mathrm{SB}} \Delta \Omega_k R_{\mathrm{ph}}^{2}}\right)^{1 / 4},
\end{equation}
where $\sigma_{\mathrm{SB}}$ is the Stephan-Boltzmann constant. The photosphere radius $R_{\mathrm{ph}}$ is determined by setting $\tau_{\mathrm{ph}}=\int_{R_{\mathrm{ph}}}^{R_{\max }} \kappa\rho(R) d R=1$ for the case $\tau_{\mathrm{tot}} > 1$. However, if the total optical depth of the ejecta $\tau_{\mathrm{tot}} \le 1$, the radius of the photosphere $R_{\mathrm{ph}}$ is taken as the minimum radius $R_{\mathrm{min}}$ \citep[e.g.,][]{Yu2018,Ren2019,Qi2022}. The polar character of this ejection allows the substance to increase its electron fraction $Y_\mathrm{e}$, inhibiting the nucleosynthesis of the heaviest $r$-process elements, and thus we use this to assign opacity. For simplicity, we set the angle to $\theta_\mathrm{b} = 30^{\circ}$ (as shown in Figure \ref{fig:kilonova}) according to \cite{Perego2017}. When in the polar direction $\theta < \theta_\mathrm{b}$, then we set $\kappa(\theta < \theta_\mathrm{b}) = 0.1\ \mathrm{{cm}^{2}{~g}^{-1}}$ and for equator direction, $\kappa(\theta > \theta_\mathrm{b}) = 10\ \mathrm{{cm}^{2}{~g}^{-1}}$. It is worth mentioning that the photon energy of the MeV range after the $r$-process are close to the gamma-ray photon energy that released from the accretion disk \citep[e.g.,][]{Metzger2019}. Thus we use the identical set of opacities for our calculation.

Then the flux density of the kilonova emission from solid angle with photon frequency $\nu$ can be given by
\begin{equation}
d F_{\nu}\left(\nu, t_{\mathrm{obs}}\right)=\frac{2 \pi h \nu^{3}}{c^{2}} \frac{1}{1-\exp \left(h \nu / k_{\mathrm{B}} T_{\text {eff }}\right)} \frac{R_{\mathrm{ph}}^{2} d \Omega}{4 \pi D_{{L}}^{2}},
\end{equation}
where  $D_L$ is the luminosity distance.

Finally, we consider the isochronous surface. Tracing the travel of light, if a photon is emitted at time $t$, it will be observed at time \citep[e.g.,][]{Qi2022}
\begin{equation}
t_{\mathrm{obs}}=t+\frac{\left[R_{\mathrm{ph}}\left(\theta_{\mathrm{obs}}\right)-R_{\mathrm{ph}}(\theta)\right] \cos \Delta \theta}{c},
\end{equation}
where $\Delta \theta$ is the angle between the direction of movement of the point on photosphere and observational direction. As a result, in the first quadrant for a given angle $\theta_\mathrm{obs}$, the observed flux density of the kilonova emission can be derived by integrating (sum of the contribution of observable blocks in two hemispheres),
\begin{equation}
F_{\nu}\left(\nu, t_{\mathrm{obs}}\right)=2 \int_{0}^{\theta_{\mathrm{obs}}+\frac{\pi}{2}} \int_{0}^{\varphi(\theta)} d F_{\nu}.
\end{equation}
If the observer is located at $\varphi = 0^{\circ}$ , the range of longitudes visible to the observer in a given $\theta$ is $[-\varphi(\theta), \varphi(\theta)]$. According to the half-day arc equation,
\begin{equation}
\varphi(\theta)=\arccos \left(-\cot \theta \cot \theta_{\text {obs }}\right).
\end{equation}

Here we take $D_L=40\ \mathrm{Mpc}$ as the distance from the source and AB magnitude $M_\nu$ can be calculated from the flux density as $M_{\nu}=-2.5 \log _{10}\left(F_{\nu} / 3631 \mathrm{Jy}\right)$.

\subsection{MeV neutrinos}

In the NDAF model, the cooling process of neutrinos occurs in large quantities, and mainly the inner region of the disk \citep[e.g.,][]{Liu2016,Wei2019,Song2020,Qi2022}. As with gamma-ray photons, the general relativity effects from the central BH, also affects the formation of neutrino spectra. So we use the same approach, i.e., the ray tracing method \citep[e.g.,][]{Fanton1997,Liu2016}. Numerically, for every pixel of our observed image, the position of the emission source in the accretion disk can be traced back. For simplicity, assuming that the neutrinos are emitted isotropically at each radius from the equatorial plane, i.e., $i_{\mathrm{em}}=\pi / 2$, and that the disk is in Keplerian rotation; at last, by ignoring the shading effect caused by the thickness of the disk, from which the trajectory of these emitted neutrinos should satisfy the geodesic equation \citep{Carter1968}, i.e.,
\begin{equation}
\pm \int_{R_{\mathrm{em}}}^{\infty} \frac{d R}{\sqrt{l(R)}}=\pm \int_{i_{\mathrm{em}}}^{i_{\mathrm{obs}}} \frac{d i}{\sqrt{I(i)}}.
\end{equation}

The energy shift of neutrinos can be calculated by considering the corresponding velocity and the gravitational potential of the emission location. The total observed spectrum is obtained by integrating all the pixels. Specifically, the total observed flux can be expressed as
\begin{equation}
F_{E_{\mathrm{obs}}}=\int_{\text {image }} g^{3} I_{E_{\mathrm{em}}} d \Omega_{\mathrm{obs}},
\end{equation}
where $E_{\mathrm{em}}$ and $E_{\mathrm{obs}}$ are the neutrino emission energy from the local disk and the observed neutrino energy, respectively. $g \equiv E_{\mathrm{obs}} / E_{\mathrm{em}}$ is the energy shift factor, and $\Omega_{\mathrm{obs}}$ is the solid angle of the disk image towards the observer.

The local emissivity $I_{E_{\mathrm{em}}}$ can be obtained from cooling rate of electron antineutrinos $Q_{\bar{\nu}_{e}}$, i.e.,
\begin{equation}
I_{E_{\mathrm{em}}}=Q_{\bar{\nu}_{e}} \frac{F_{E_{\mathrm{em}}}}{\int F_{E_{\mathrm{em}}} d E_{\mathrm{em}}},
\end{equation}
where $F_{E_{\mathrm{em}}}=E_{\mathrm{em}}^{2} /\left[\exp \left(E_{\mathrm{em}} / k T-\eta\right)+1\right]$ is the unnormalized Fermi-Dirac spectrum \citep[e.g.,][]{Liu2016,Wei2019}.

Hence, the luminosity distribution can be calculated as follows
\begin{equation}
L_{\nu}=4 \pi D_L^2 F_{E_{\mathrm{obs}}}.
\end{equation}
It should be mentioned that in addition to the neutrinos released from the NDAFs, the vertical advection mechanism also releases a very small fraction of neutrinos.

\subsection{GWs}

\begin{figure}
\centering
\includegraphics[width=0.5\textwidth]{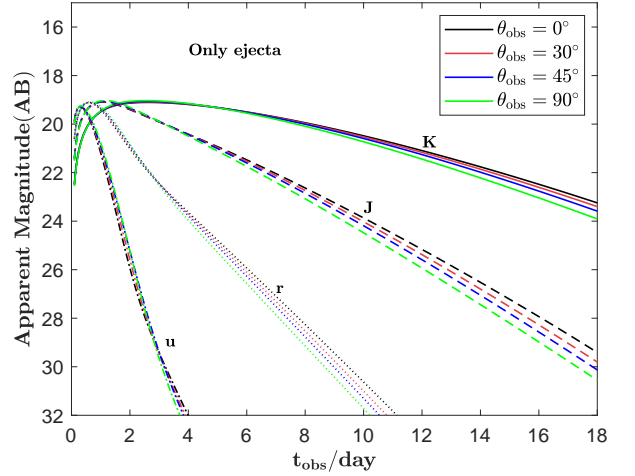}
\caption{urJK-band light curves of ejecta-driven kilonova at a distance of 40 Mpc. The black, red, blue, and green lines indicate the observation angles $\theta_{\mathrm{obs}}=0^{\circ}, 30^{\circ}, 45^{\circ}$, and $90^{\circ}$, respectively. The solid, dashed, dotted, and dash-dot lines represent the $\mathrm{K}, \mathrm{J}, \mathrm{r}$, and u bands, respectively.}
\label{fig:ejecta}
\end{figure}

GW radiation induced by the anisotropic neutrino emission was first investigated by \citet{Epstein1978} and this approach has been applied to core-collapse supernovae \citep[e.g.,][]{Burrows1996, Kotake2006, Kotake2007} and NDAFs \citep[e.g.,][]{Suwa2009,Kotake2012,Liu2017b,Song2020,Wei2020}. Here, we adopt the method to calculate the GWs from NDAFs with vertical advection in compact binary merger scenarios.

For long-term neutrino emission, the GW amplitude will converge to a nonzero value $h_{\infty}$ and it is subject to the observation angle $\theta_{\mathrm{obs}}$, which is derived as follows \citep{Suwa2009}
\begin{equation}
h_{\infty}\left(\theta_{\mathrm{obs}}\right)=\frac{2 G\left(1+2 \cos \theta_{\mathrm{obs}}\right)}{3 c^{4} D_{{L}}} \tan ^{2}\left(\frac{\theta_{\mathrm{obs}}}{2}\right) \bar{L}_{\nu} \tilde{T},
\end{equation}
where $\bar{L}_{\nu}=2 \pi \int_{0}^{\tilde{T}} \int_{r_{\text {inner }}}^{r_{\text {outer }}} Q_{\nu}^{-} R\ d R\ d t / \tilde{T}$ is the mean neutrino luminosity above or below the disk, $\tilde{T}$ is the activity duration of the GRB central engine. If the GRB is considered as a single burst triggered by the NDAF, the time evolution of the neutrino luminosity $L_{\nu}(t)$ can then be described as
\begin{equation}
L_{\nu}(t)=\bar{L}_{\nu} \Theta(t) \Theta(\tilde{T}-t),
\end{equation}
where $\Theta$ is the Heaviside step function. For the case of a possibly more realistic multiple pulses, it can be expressed as
\begin{equation}
L_{\nu}(t)=\sum_{i'=1}^{N^{\prime}} \frac{\bar{L}_{\nu} \tilde{T}}{N^{\prime} \delta t} \Theta\left(t-\frac{i'}{N^{\prime}} \tilde{T}\right) \Theta\left(\frac{i'}{N^{\prime}} \tilde{T}+\delta t-t\right),
\end{equation}
where $N^{\prime}$ is the number of bursts and $\delta t$ is the duration of one burst. $N^{\prime} \delta t$ should be shorter than total duration $\tilde{T}$, unless the case of a single pulse of $N^{\prime}=1$.

\begin{figure*}
\centering
\includegraphics[width=0.49\textwidth]{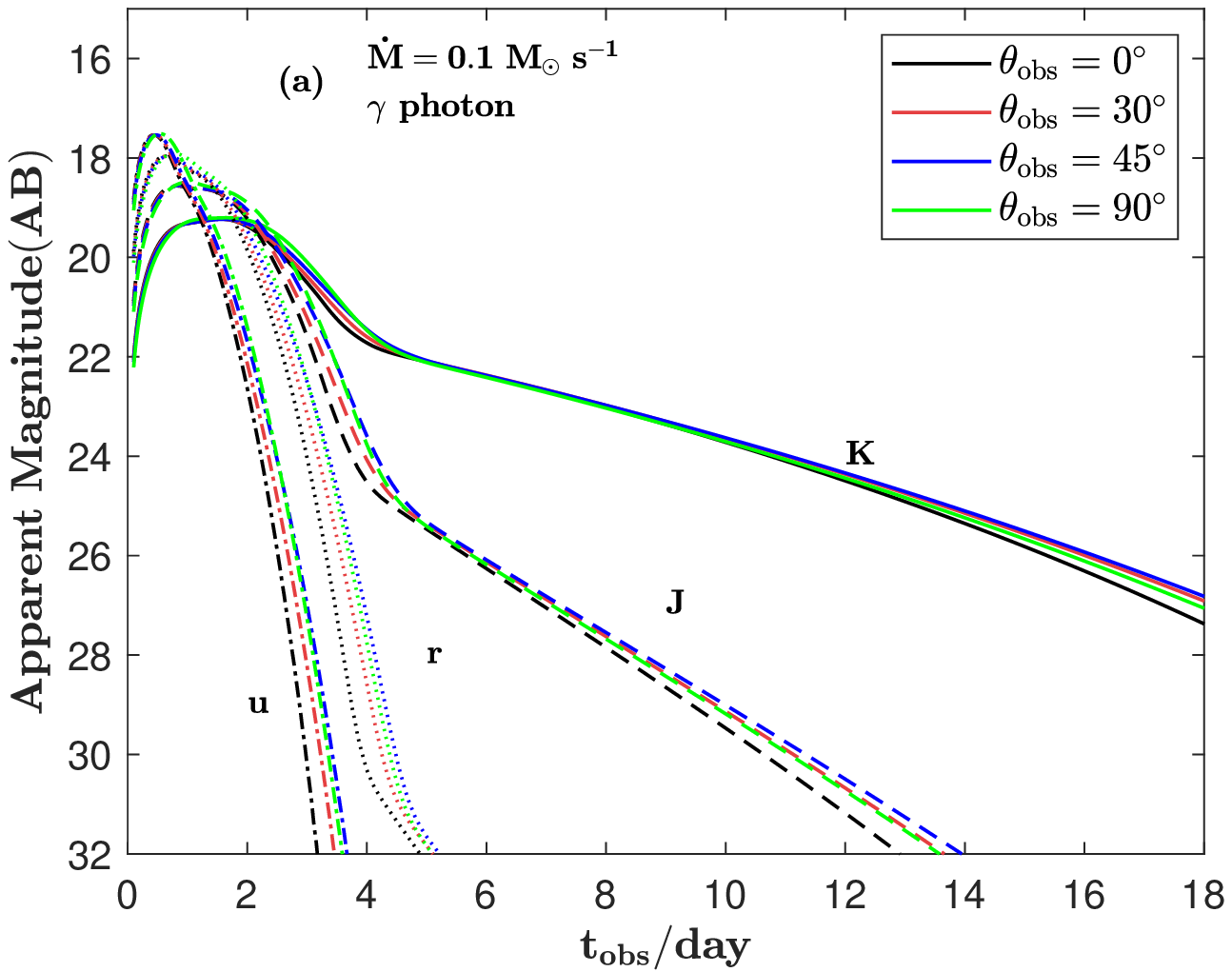}
\includegraphics[width=0.49\textwidth]{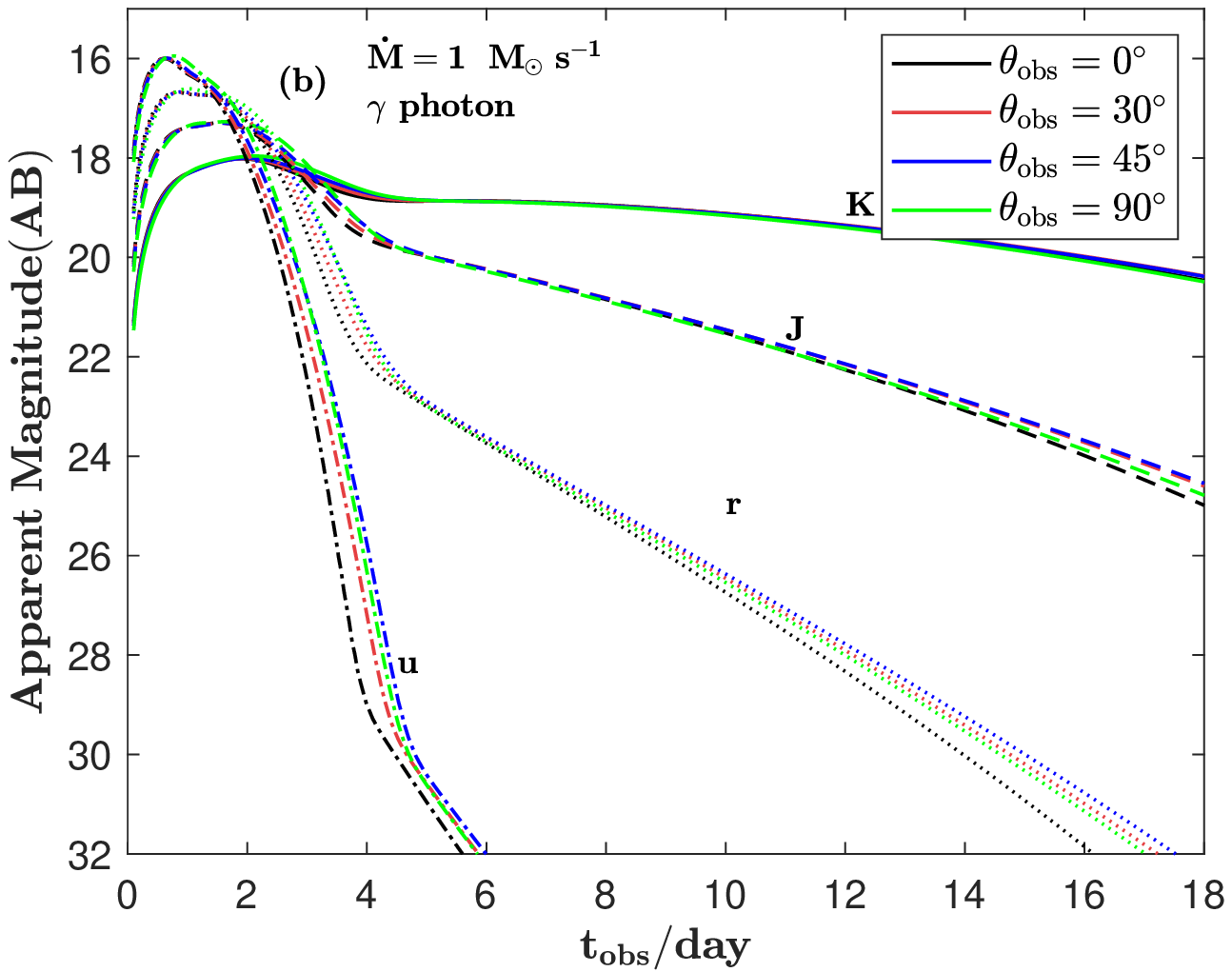}
\includegraphics[width=0.49\textwidth]{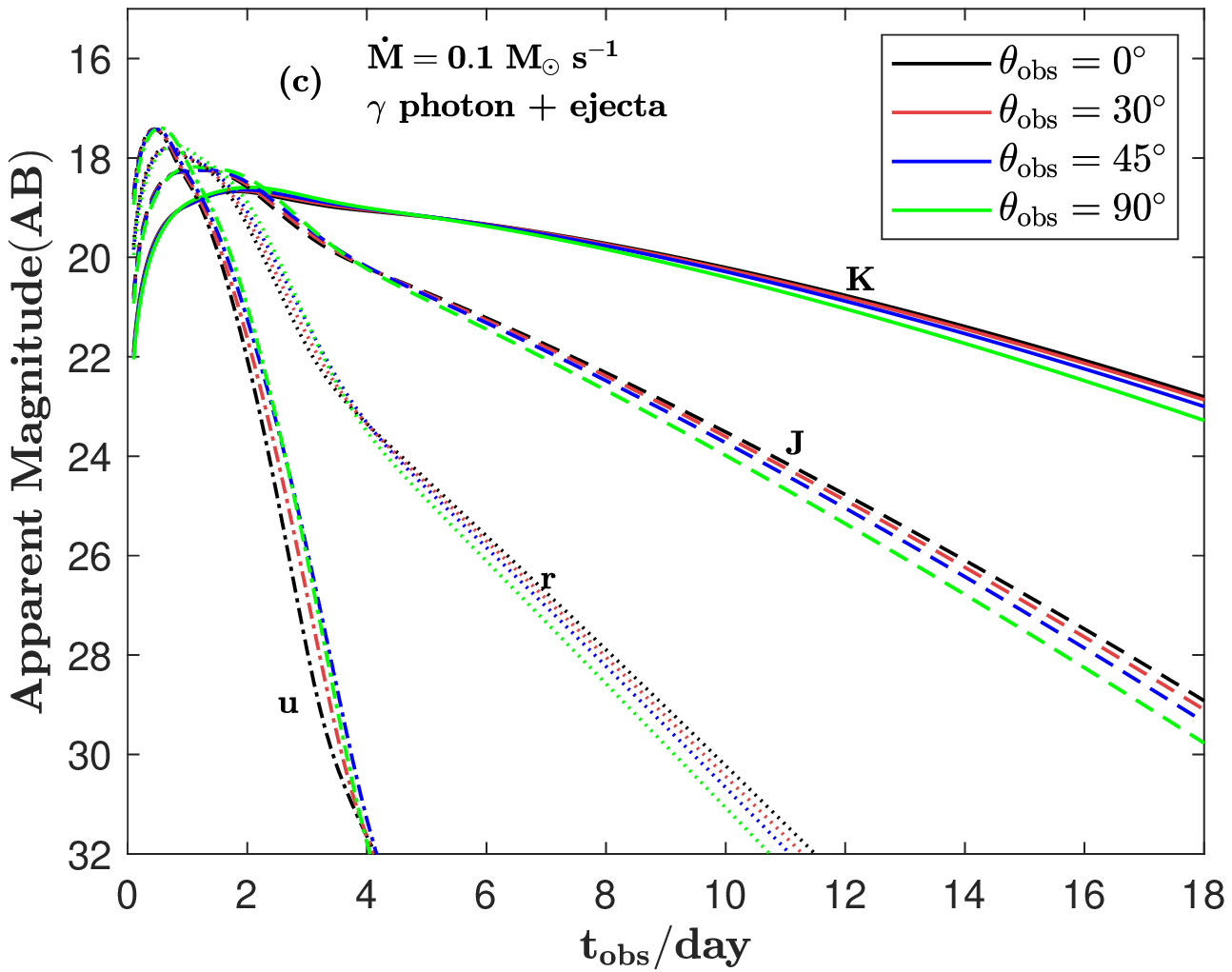}
\includegraphics[width=0.49\textwidth]{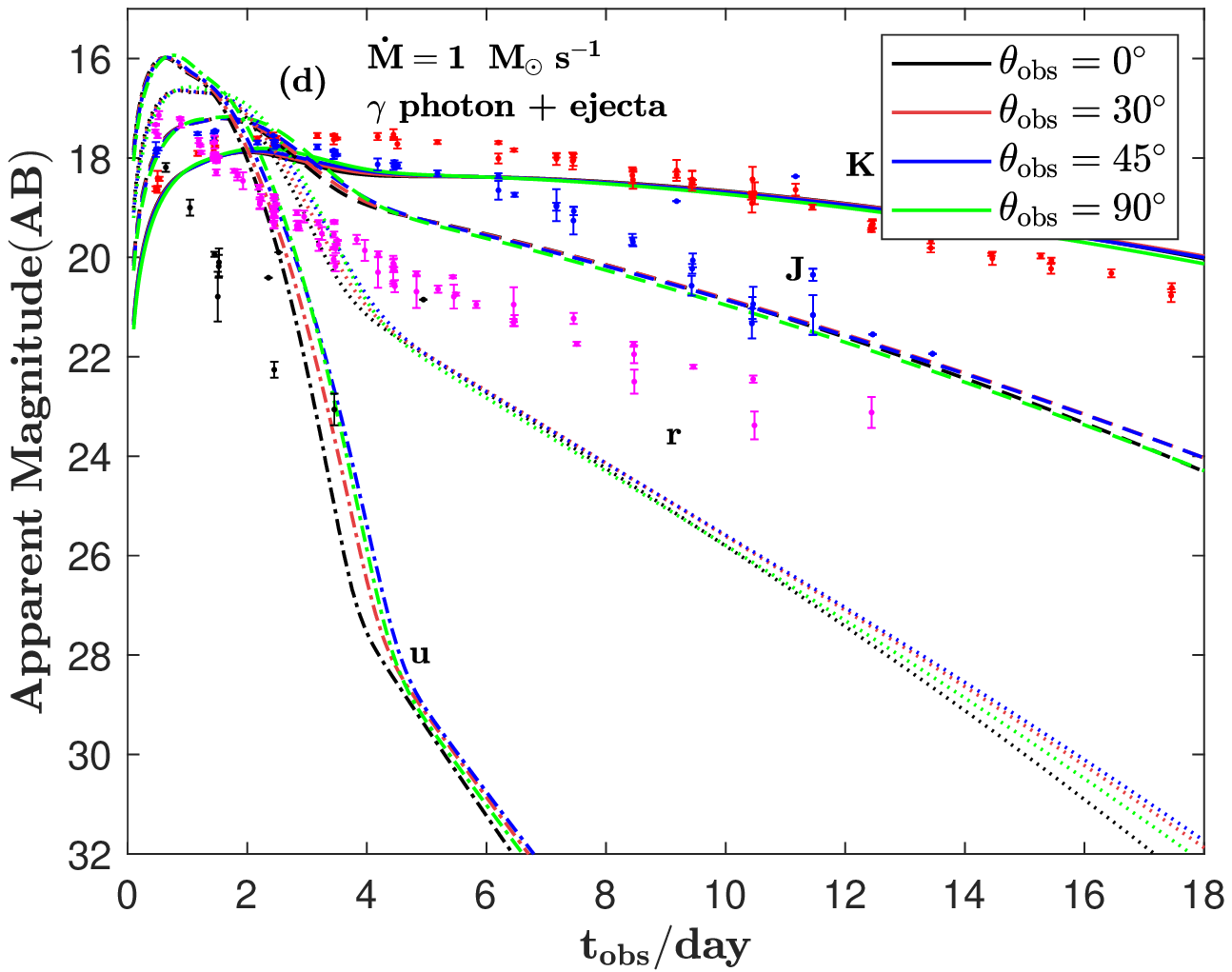}
\caption{Viewing-angle-dependent urJK-band light curves of kilonovae at a distance of 40 Mpc for the cases of kilonovae driven by gamma-ray photons only and by ejecta radioactive decay and gamma-ray photons together at an accretion rate of $0.1$ or $1\ M_{\odot}\ \mathrm{s}^{-1}$. The black, red, blue, and green lines indicate the observation angles $\theta_{\mathrm{obs}}=0^{\circ}, 30^{\circ}, 45^{\circ}$, and $90^{\circ}$, respectively. The solid, dashed, dotted, and dash-dot lines represent the $\mathrm{K}, \mathrm{J}, \mathrm{r}$, and u bands, respectively. Panel (d): comparison between the urJK-band light curves for AT 2017gfo and our kilonova model. The data are taken from \cite{Villar2017}. The red, blue, magenta and black dots represent the K, J, r and u bands respectively.}
\label{fig:photons}
\vspace{0.2in}
\end{figure*}

After the Fourier inversion, the form of $L_{\nu}(t)$ can be given as
\begin{equation}
L_{\nu}(t)=\int_{-\infty}^{\infty} \tilde{L}_{\nu}(f) e^{-2 \pi i f t} d f,
\end{equation}
where $f$ is the frequency.

Considering an axisymmetric sources, the only nonvanishing component of the GW amplitude of the NDAF is \citep[see more detalis in][]{Mueller1997}
\begin{equation}
h_{+}(t)=\frac{2 G}{3 D_{L} c^{4}} \int_{-\infty}^{t-D_{L} / c} L_{\nu}\left(t^{\prime}\right) d t^{\prime}.
\end{equation}

Therefore, the local energy flux of GWs can be expressed as \citep[e.g.,][]{Suwa2009,Wei2020}
\begin{equation}
\frac{d E_{\mathrm{GW}}}{D_{L}^{2} d \Omega d t}=\frac{c^{3}}{16 \pi G}\left|\frac{d}{d t} h_{+}(t)\right|^{2}.
\end{equation}

The total GW energy can be written as
\begin{equation}
E_{\mathrm{GW}}=\frac{\beta G}{9 c^{5}} \int_{-\infty}^{\infty} L_{\nu}(t)^{2} d t,
\end{equation}
with $\beta \sim 0.47039$. After deriving the previous equations, we can readily obtain the GW energy spectrum for NDAF model which is
\begin{equation}
\frac{d E_{\mathrm{GW}}(f)}{d f}=\frac{2 \beta G}{9 c^{5}}\left|\tilde{L}_{\nu}(f)\right|^{2} .
\end{equation}

For the goal of discussing the detectability of GW, we write down the characteristic GW strain as
\begin{equation}
h_{c}(f)=\sqrt{\frac{2}{\pi^{2}} \frac{G}{c^{3}} \frac{1}{D_{L}^{2}} \frac{d E_{\mathrm{GW}}(f)}{d f}}.
\end{equation}

Finally, the signal-to-noise ratios (S/Ns) obtained through matched filtering from the GW experiments can also be derived, the optimal S/N is
\begin{equation}
\mathrm{S} / \mathrm{N}^{2}=\int_{0}^{\infty} \frac{h_{c}^{2}(f)}{h_{n}^{2}(f)} \frac{d f}{f},
\end{equation}
where $h_{n}(f)=\left[5 f S_{h}(f)\right]^{1 / 2}$ is the noise amplitude with $S_{h}(f)$ being the spectral density of the strain noise.

\section{Results}

\subsection{Disk Structure}

The calculation results show that the vertical advection process plays an important role on the physical properties of the disk. By solving the above Equations (\ref{NDAFfirst}) $-$ (15) for NDAFs, we present the radial profiles of the disk density and temperature for both cases with and without vertical advection in Figure \ref{fig:NDAFsolution}. Two typical accretion rates $\dot{M} = 0.1$ and $1\ M_{\odot}\ \mathrm{s^{-1}}$ are adopted in the graph by red and blue lines, respectively. It is explicitly to note that when the accretion rate is at a different order of magnitude, there is a considerably differences in properties of the disk. For a given accretion rate and radius, one with vertical advection will have higher density and lower temperature compared with the other.

The physical explanation is that through vertical advection process a large quantity of trapped photons are radiated from the originally optically thick hyper-accretion disk due to the magnetic buoyancy. The energy is carried out by these gamma-ray photons, resulting a lower temperature and lower radiation pressure, but higher disk density. It is worth noting that such significant gamma-ray photons should be detectable, or it may inject into the electromagnetic counterpart in binary compact object merger events, contributing to the luminosity of kilonovae.

\subsection{Kilonovae}

\begin{figure*}
\centering
\includegraphics[width=0.49\textwidth]{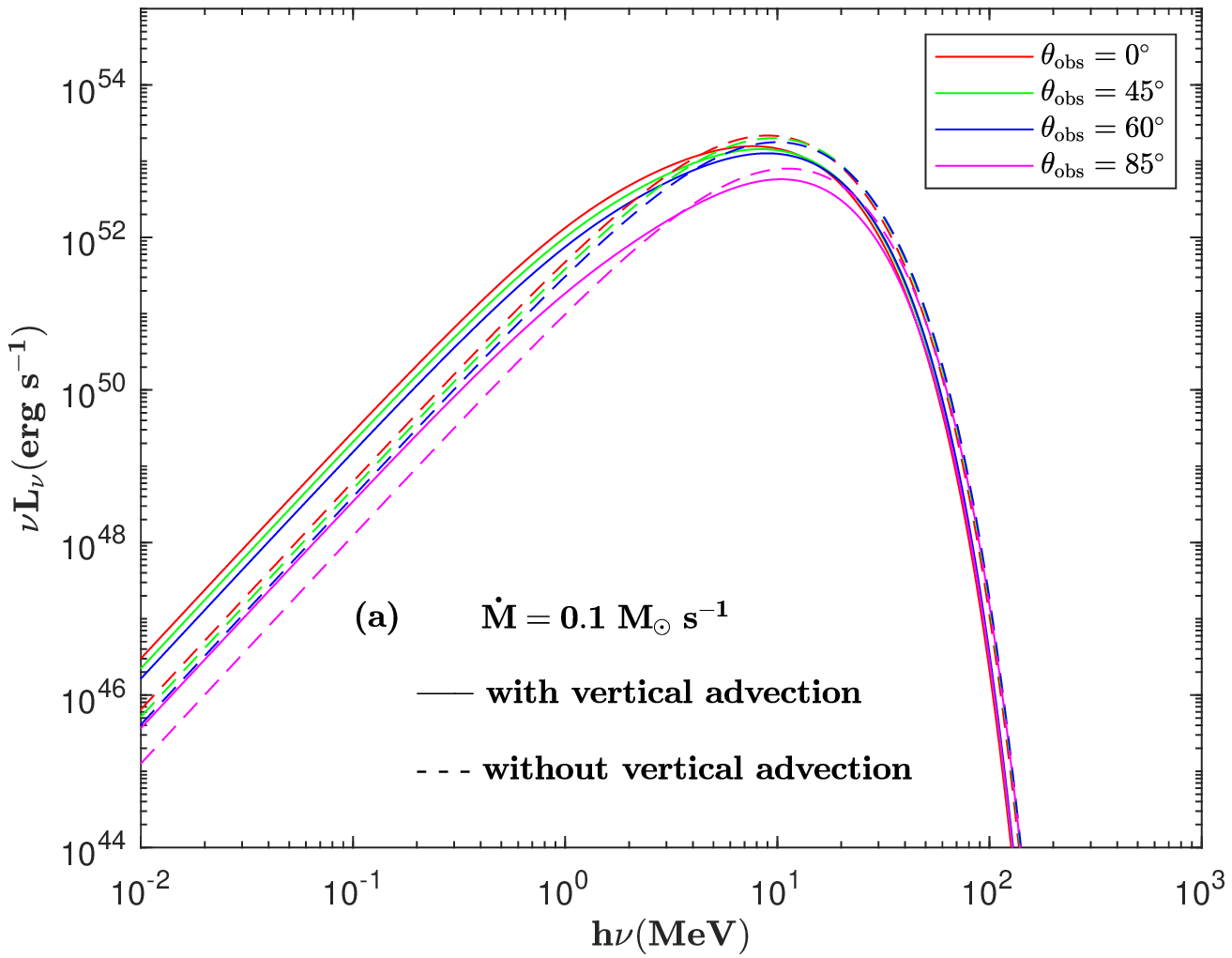}
\includegraphics[width=0.49\textwidth]{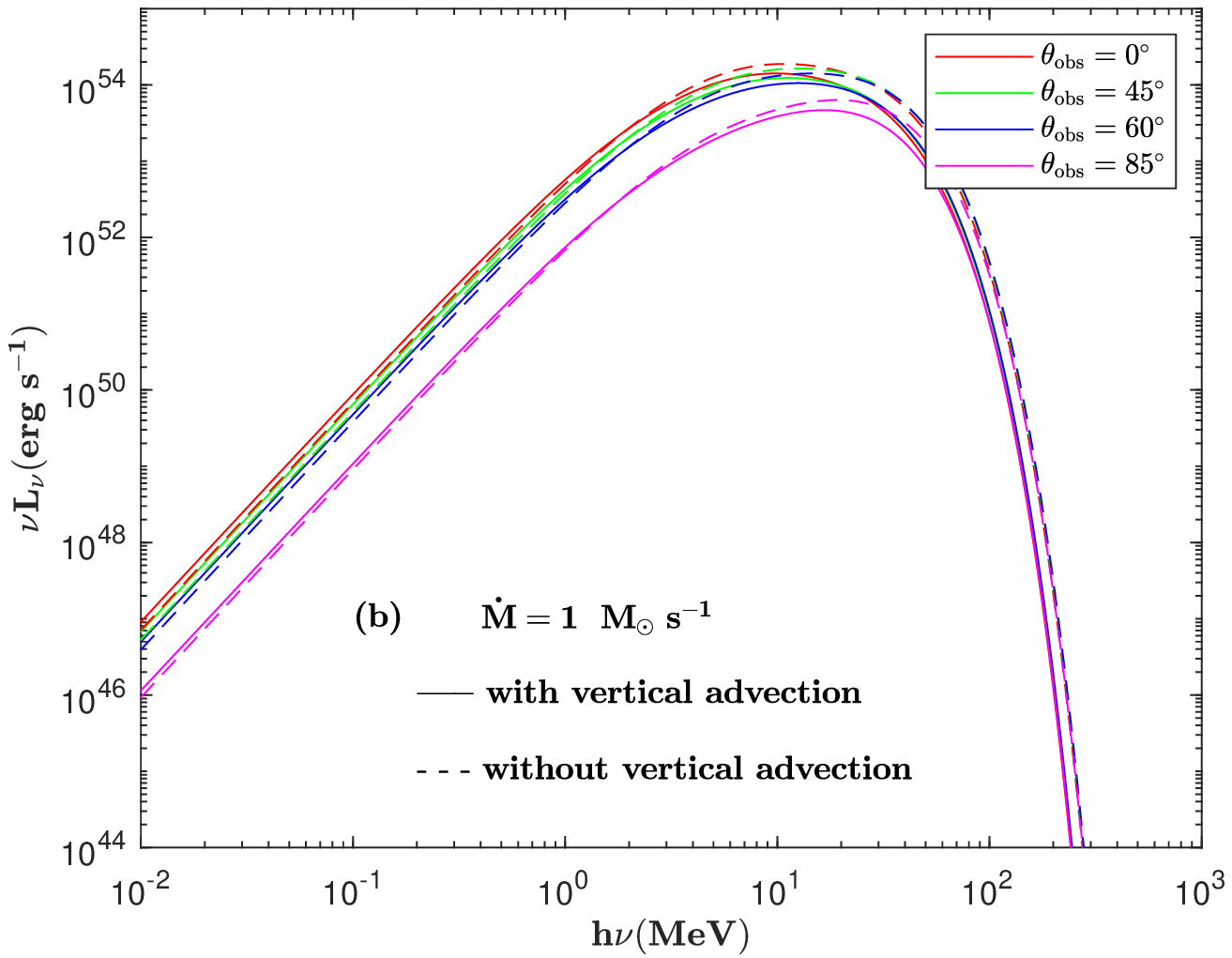}
\caption{Electron neutrino and antineutrino spectra as functions of viewing angle of NDAFs with (solid lines) and without (dashed lines) vertical advection in merger scenario. The red, green, blue, and magenta lines correspond $\theta_{\mathrm{obs}}=0^{\circ}, 30^{\circ}, 45^{\circ}$, and $85^{\circ}$, respectively.}
\label{fig:neutrinos}
\vspace{0.2in}
\end{figure*}

The anisotropic emission resulting from the energy injection differences of gamma-ray photons at different angles is the main focus in the present work, but first we enumerate the case where there is no energy injection, i.e., kilonova triggered by the radioactive decay of ejecta alone.

We show the viewing-angle-dependent urJK-band light curves of pure ejecta-driven kilonova in Figure \ref{fig:ejecta} at a distance of 40 Mpc. The black, red, blue, and green lines indicate the observation angles $\theta_{\text {obs }}=0^{\circ}, 30^{\circ}, 45^{\circ}$, and $90^{\circ}$, respectively. The K, J, r, and u bands are represented by the solid, dashed, dotted, and dash-dot line, respectively. As we can see that, the kilonova luminosity has the typical magnitudes of no more than $\sim 19$ mag and each band decrease correspondingly with the decrease in observation angle $\theta_\mathrm{obs}$. That is because the mass and velocity of the ejecta is changing with the distribution of the angle. In fact, the biggest effect to the anisotropy is due to the difference in opacity at $\theta_\mathrm{b}$ as a boundary, i.e., the red and blue components of the kilonova, because the differences in angle-dependent opacity amplify the contribution to anisotropy. The peak is caused by the BC produced by low-opacity ejecta near the polar axis.

Next, we compared in Figure \ref{fig:photons} the four cases of the presence of energy injection, i.e., kilonovae only driven by gamma-ray photons and by ejecta radioactive decay and gamma-ray photons together at an accretion rate of $0.1$ or $1\ M_{\odot}\ \mathrm{s}^{-1}$, respectively. One can find that a distinct common feature is that at the very early phase of the light curve there is a sharp bulge covering the original ejecta curve, where the luminosity increases dramatically and peaks at $\sim 1$ day. With an accretion rate of $0.1\ M_{\odot}\ \mathrm{s^{-1}}$, the peak magnitudes increase to over $\sim 18$ mag in the ultraviolet band and over $\sim 19$ mag in the infrared band. However, at an accretion rate of $1\ M_{\odot}\ \mathrm{s^{-1}}$, these values increase to $16$ mag and $18$ mag, respectively. After that, the luminosity curve decays rapidly and then goes into a smooth declining phase at $\sim 4$ day. Actually, comparing between (a) and (c) or (b) and (d), the gamma-ray photons contribute throughout the whole light curves. It is worth noting that at higher accretion rates, the peak becomes more apparent. That is because the appearance of peak structure in the light curves is related to the highly injected energy of gamma-ray photons that released from the accretion disk is significantly high. In the early stage, the optical depth is high, the gamma-ray photons are mainly absorbed by the inner region of the ejecta, and the radiative cooling of the ejecta occurs mainly in the outer region. That is, the early energy injection was mainly deposited in the inner layer. As the optical depth $\tau_\mathrm{tot}$ decreases, these deposited energies will be released and its value is associated with the size of the bulge. By comparing the images longitudinally with an accretion rate of $0.1\ M_{\odot}\ \mathrm{s^{-1}}$, we find the consistency of the early curves. The luminosity driven by the gamma-ray photons fades rapidly after $\sim$ 3 days. Instead, in the later phase, the effect of radioactive decay emission causes by the ejecta emerges and dominates over the emission. The transition thus causes an ``ankle'' at $\sim$ 3 days, as seen in the light curves, with later stages decaying slower than the earlier stage. This time scale corresponds to the diffusion time scale of the ejecta. After the transient injection, the energy needs to pass through layers of ejecta, from the innermost layer to the outermost layer, which is essentially determined by the mass and velocity of the ejecta. If we compare the case of an accretion rate of $1\ M_{\odot}\ \mathrm{s^{-1}}$, the energy of the radioactive decay is less important, the luminosity of the kilonova is increased considerably, and the whole process is almost dominated by the injected energy. The observed radiation would be mainly from the gamma-ray photons. The trajectory of the photon changes due to the gravitational force of the BH, so the magnitude of the injected energy varies with the angle, which still shows anisotropy.

Figure \ref{fig:photons}(d) also shows the comparison between the light curve of AT2017gfo and our kilonova model with $\dot{M} = 1\ M_{\odot}~\rm s^{-1}$. One can notice that in the early phase, the light curve is higher than the fitting point, which is probably due to the overestimation on the efficiency of the transient injected energy or the production of the vertical advection cooling rates.

\begin{figure*}
\centering
\includegraphics[width=0.49\textwidth]{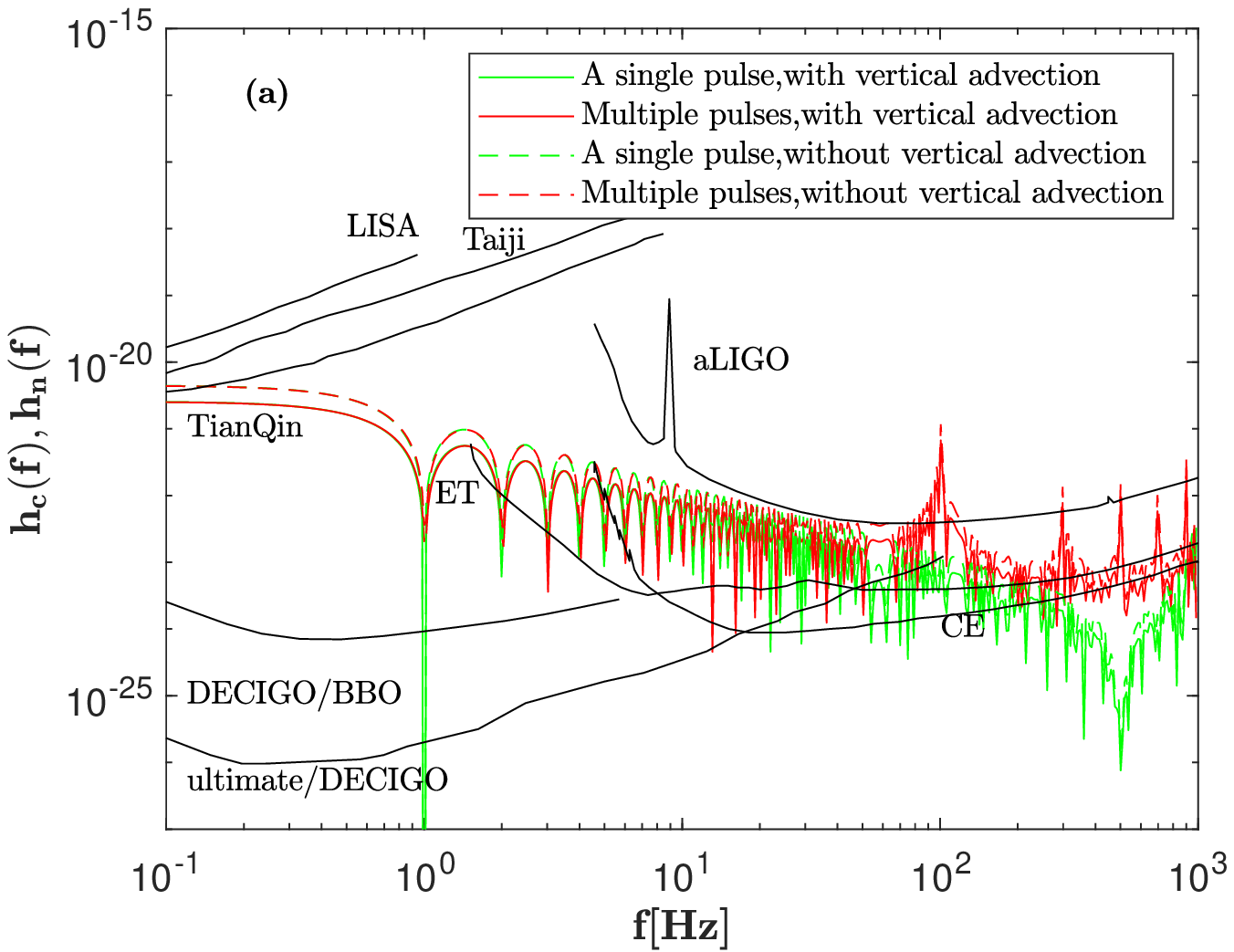}
\includegraphics[width=0.49\textwidth]{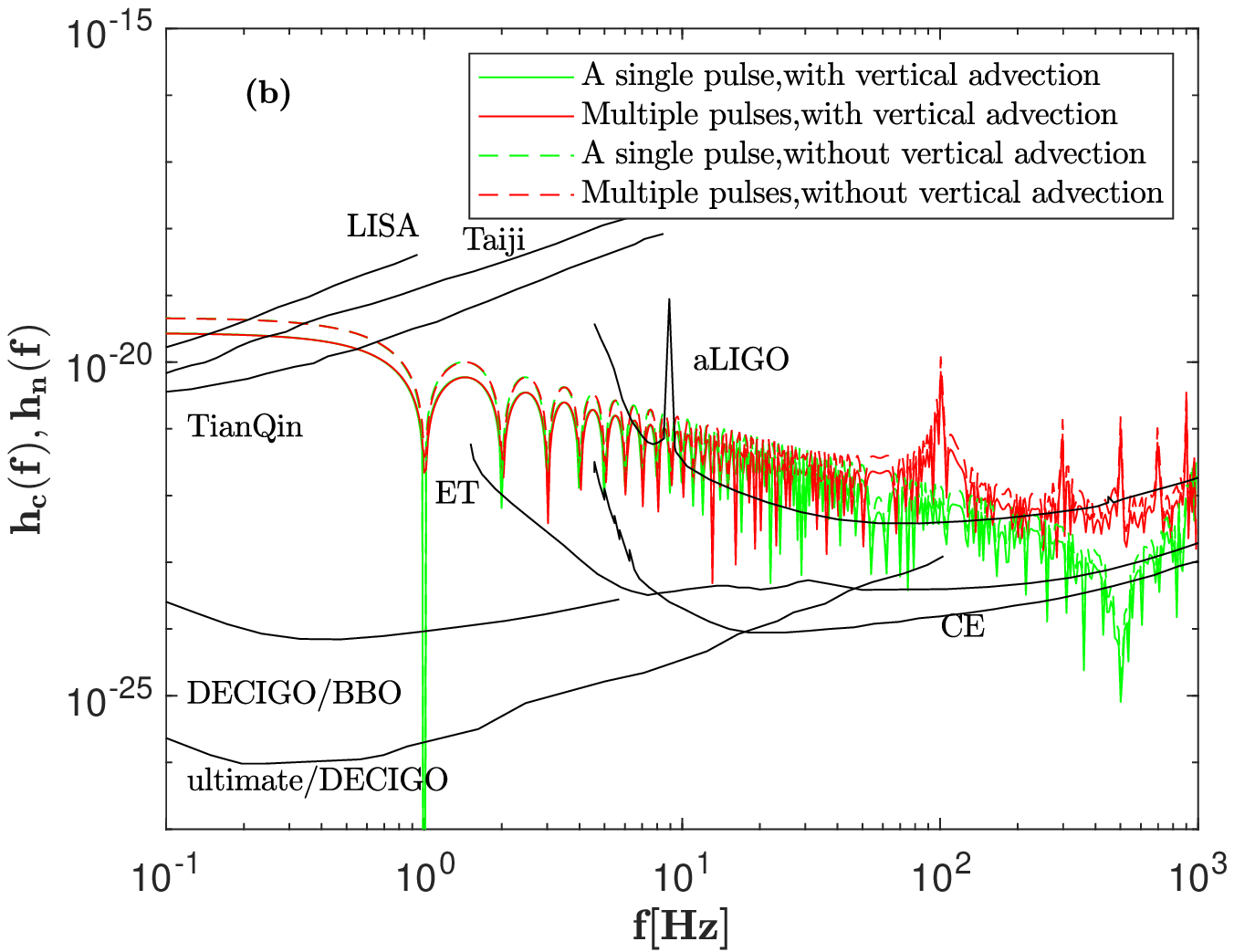}
\caption{The strains of GWs from NDAFs with (solid lines) or without (dashed lines) vertical advection in merger scenarios at a distance of 10 kpc. The green and red lines represent the cases of SGRBs with a single pulse and the multiple pulses, respectively. The black solid lines represent the sensitivity lines (the noise amplitudes $h_{\mathrm{n}}$) of aLIGO, ET, CE, LISA, Taiji, TianQin, DECIGO/BBO, and ultimate-DECIGO.}
\label{fig:GW}
\vspace{0.2in}
\end{figure*}

\subsection{Neutrino emission}

The magnetic buoyancy produced by the vertical advection effect also leads to the escape of MeV neutrinos. We forecast the expected neutrinos anisotropic radiation from the NDAFs model, the neutrino energy is normally in the range of $\sim 1-$ 100 MeV, with a peak ranging from $10 - 20$ MeV. Figure \ref{fig:neutrinos} shows the electron antineutrino spectra as a function of viewing angle of NDAFs with (solid lines) or without (dashed lines) vertical advection. The red, green, blue, and magenta lines correspond to  $\theta_{\mathrm{obs}}=0^{\circ}, 30^{\circ}, 45^{\circ}$, and $85^{\circ}$, respectively. The energy of neutrinos is greater at higher accretion rates. The contribution of the vertical advection in NDAFs to neutrino yields is very limited. In the absence of vertical advection, the peak of neutrino is slightly shifted toward greater energy. In the high-energy range of the spectra, the neutrinos with the vertical advection have a lower luminosity. This is due to the fact that neutrino emission is very sensitive to temperature, and like the results shown in Figure \ref{fig:NDAFsolution}, the disk temperature is lower in the case with vertical advection, which leads to a lower reaction rate for generating neutrinos, even though a fraction of neutrinos are released by the bubbles. As a result, the neutrino energy is lower than in the case without vertical advection. Note that the optical depth of neutrinos is much smaller than that of the photons, the bubble actually cannot effectively confine many neutrinos inside. Therefore, the neutrinos are affected by the vertical advection effect but only slightly, as compared to the photons. Besides, it is clear that the low-energy $(\lesssim 10\ \mathrm{MeV})$ neutrinos have a more anisotropic luminosity and the amplitudes of the spectral lines decrease with increasing observation angels.

\subsection{GWs}

Figure \ref{fig:GW} shows the strains of GWs from NDAFs with or without vertical advection at a distance of $10\ \mathrm{kpc}$ for cases of SGRBs with a single pulse ($\tilde{T} = 1\mathrm{~s}$) and the multiple pulses ($\tilde{T} = 1\mathrm{~s}$, $N^{\prime}=100$, and $\delta t=0.001 \mathrm{~s}$) for the accretion rates $0.1$ and $1\ M_{\odot}\ \mathrm{s^{-1}}$ (corresponding Panels (a) and (b)). The black solid lines are the sensitivity curves (the noise amplitudes $h_{n}$) of aLIGO, Einstein Telescope (ET), Cosmic Explorer (CE), Laser Interferometer Space Antenna (LISA), Taiji, TianQin, Decihertz Interferometer Gravitational Wave Observatory/Big Bang Observer (DECIGO/BBO), and ultimate-DECIGO detectors. Comparing the cases with different accretion rates, when the accretion rate is higher, the more anisotropic neutrinos are produced on the accretion disk and the GW strain increases by several orders of magnitude. However, for the cases with same accretion rates, the GW strain increases slightly without the vertical advection effect.

The typical GW frequencies are noted to be in the range of 1 $-$ 1,000 Hz, which is determined by the variability of the GRB. The GW waveforms caused by a single pulse and multiple pulses are quite different, which also results from the difference in the variability of neutrino emission. It can be seen that the GWs (or memory) induced by the NDAF with vertical advection can be detected by ET, CE, DECIGO/BBO, and ultimate-DECIGO. This is essential to constrain the nature of the central engine as well as the neutrino radiation. Since the GW strains depend entirely on the neutrino luminosity and the very limited contribution of the vertical advection to neutrino yields (also see Figure 6), there is tiny difference on the strains of GWs from NDAFs between with and without vertical advection.

\section{SUMMARY}

In the scenario of compact binary coalescences, a BH hyperaccrection disk is formed and which thought to be an NDAF because of the extreme physical conditions triggering the neutrino radiation. In particular, during the coalescences of BH-NS or NS-NS, it is possible to detect multimessenger signal emission, such as SGRBs, kilonovae, neutrinos, and GWs, for which a theoretical prediction can be made and allows us to have a more comprehensive understanding of the source. In this paper, we model the NDAFs with additional vertical advection process and introduce the relativistic correction factors to correct the effect of BH spin on the disk structure. We found that the vertical advection process slightly increases the disk density while decreasing the temperature. We present and analyze multi-band (utJK) kilonovae light curves derived from our model. We noticed that the different masses, velocities, and opacity distributions of the ejecta are responsible for the anisotropy of the kilonovae, but the opacity causes the most significant effect. In addition, the gamma-ray photons escaping from the NDAF with vertical advection radiate anisotropically due to the gravitational influence of the BH and this considerable energy is injected into the ejecta, resulting in a significant increase in the luminosity of the early phase of the kilonovae. Then we study the electron antineutrino energy spectra from NDAFs with and without vertical advection at different observation angles and observed more anisotropic radiation in the low-energy bands. In the future neutrino detectors may have the ability to provide valuable detections of neutrinos and even their spectra to verify the existence of NDAFs in the center of compact binary coalescences or massive collapsars. Finally, we also present the GW emission induced by the anisotropic neutrino radiation of NDAFs with vertical advection. The GW strain is completely determined by the neutrino luminosity and the GW signal from NDAFs with or without vertical advection at a distance of 10 kpc can be detected by CE, ET, ultimate DECIGO, and DECIGO/BBO. Although there are slight differences on neutrino spectra and GW shapes of NDAFs between with and without vertical advection, the future joint multimessenger observations might distinguish them for the sources in the local universe.

\acknowledgments
This work was supported by the National Natural Science Foundation of China under grants 12173031 and 12221003.

\end{document}